# Digital Twins: How Far from Ideas to Twins?

*Lu Jingyu*

*Department of Mechanical Engineering, Jiangnan University, Wuxi, China*

*7200832004@stu.jiangnan.edu.cn*

# Abstract

As a bridge from virtuality to reality, Digital Twin has increased in popularity since proposed. Ideas have been proposed theoretical and practical for digital twins. From theoretical perspective, digital twin is fusion of data mapping between modalities; from practical point of view, digital twin is scenario implementation based on the Internet of Things and models. From these two perspectives, we explore the researches from idea to realization of digital twins and discuss thoroughly.

# 1 Introduction

Digital twins (DT) have been booming since it was proposed by Michael in 2002[1]. Michael interprets that DT can be designed, tested, manufactured, and be used as a virtual version. Due to the ubiquitous role of the digital twin in the total product life cycle, it is oriented to complex systems[2], DT will face more possible problems more freely and vividly.

What can DT do will certainly make attention and there are many researchers considered that the DT can not only simulating but also create a twin world which can help industry to revolute a virtual copy of the real world[3]. If we dive in from this point of view, how real is it about the world that DT present to us and how can we believe it?

We start comment in two direction which are theories and practical. The writer admitted the view is limited especially for the practical. We believe that the DT have gone further than we can see. In this manuscript we only discussed the influence which we can see as far and as more as possible. There are various types of publication which has been discussed DT. From the Google scholar to WOS, we can discover that the DT has not been formally proposed until 2000 around and couple researchers has made their discussion from the security of complex control systems[4] to product quality analysis[5]. When search for the earliest essay related to DT found before 2000 there has been similar publication about the virtual environment[6] that is likely with DT in concept.

Just like BIM (Building Information Modeling), DT expend people's horizon both in theories and practical which helps engineers, researchers, and decision makers with a virtual view. In particular, DT have made a huge publication in manufacture. We search DT in WOS and classified by research area 'Engineering' is the most publication and the number of publications has been rising from 39 in 2000 to 1858 in 2022 and the citation increased from 148 in 2000 to 26111.

Take a look at it from the perspective of Engineering. The most citied paper Tao et.al[7] consider that DT has improved the production process from product design to the production and make a full life cycle management which end up to combine a cyber-physical fusion. If we take a deep look at this part. It will separate into different kinds of data. For example, if you want to make an independent component virtual what is the specific information that the user in the virtual world would like to see and which is the should one and which is the must one? Another issue to consider is the role of the user in the real world, which should be ensured that the relationship between the role and the user should be mapped correctly.

The independent components can be shown in the virtual world which can be modeled in some 3D software and then shown to users by using different programming languages. During this process, various kinds of data formats conversion occurred. This depends on the Human-Machine Interface (HMI) type. The industry use like PDA or PC then file format can be imported or designed in software like Solidworks et.al[8]. After the transition the HMI can show the result as a virtual vision of component. The role of user which depend on the industry and can be changed once the role has changed. For the transition by the manufacturing if type of the workshop is metalworking frame, the DT need to face the reality that the different type of machine tool will have different kind of PLC which has a vital role for the transition. The PLC has the information of how and when to control and change the status of tools during the production process. This process depends on the craft card and related to the actual operation of the tools.

As for the operation of machine tool, there will be numerous of sensors which help the transition successfully by transforming the signal into data. Due to the different type of sensors the transition will not exactly same. Whether sensor data or the decision data both of them all need a specific action which defined information flow[9]. Which allow the physical environment and system can undertake the actions and make the instruction passed to the specified location of the system. While try to let the machine understand the instruction and do it correctly it not that simple. Not only to make sure the instruction is correct to environment also confirm the command should do correct signal or order in specific system. However, the system may be chaos [10] or discrete[11] which means that even if the command only considers the start and terminal of the command there still tons of uncertainty in the process and we will move forward in Section 2.

Meanwhile the development of AI (Artificial Intelligence), there are technologies which help the researchers or industry to make it happen. The role of AI may change based on the role of the DT. And Deep Learning (DL) now has strong developed especially in image[12], video[13] and nature language processing[14]. Certainly, the emergence of DL enables researchers to effectively engage with various data models that may arise in the processing of DT[15].

The remainder of this paper organized as follows. In section 2 we introduce different industry which highly showed interest in DT and summary issues and methods. And then by listing some of theoretical method which has been mentioned serval time and see how does these methods affect DT. Finally in section 3 we discuss different technologies which may benefit DT.

# 2 State-of-art

## 2.1 DT in Industries

Manufacturing is one of the most popular industry that try to apply DT in different kinds of process during the manufacturing such as product design[16, 17], process monitoring[18] and quality control[19] et.al. This work is aimed to discuss the benefits and possibilities that DT can bring up to the table. Firstly, we clarify two research questions:

**RQ1:** What are major entities of DT using in these cases?
**RQ2:** What impacts have by using DT?

## 2.1.1 DT in Manufacturing

Carrying with questions, firstly we searched literatures about application of DT in manufacturing. Then we dived in and try to define the role of DT in these cases. Note that we don't dive deep into details and processes in this part or check out the theory but we offer cases to explore the impacts of DT. By using the service and process classification defined by Tao[20] and the ISO 23247-3:2021[21], classification. The entity of manufacture process and information used in literatures can be defined as Table 1.

Table1 Classification, Entity and Information used to define the DT use case.

| Process | Information | Entities | Function |
|---|---|---|---|
| Material | Material | User | Production material[22] |
| Information | Equipment | | Analyze and update |

| | | | | Operational and environmental at rate[23] |
|---|---|---|---|---|
| Energy | | Personnel | Digital twin | Monitoring energy assumption[24] during production |
| Control | Design | Process | | Conceptual and detailed design[25] to refine product scheduling[26] |
| | System error analysis and predict | Facility | Observable manufacturing elements | On-line or off-line analysis and in-loop check-in[27] |
| | Maintenance | Environment | | Health check and predictive maintenances[28] |
| | System monitor | Product | Device communication | Production management[29] |
| | System support | Supporting document | | Data simulation, security information[30], and support document based on process prediction[31] |

We reviewed the literature referred to Christos' work[32] and changing keywords into "digital twin*" AND (manufacture* OR production* OR factory*) in WOS . There are 82 highly cited papers and 18 of them are reviews. Exclude those not related to the manufacturing industry, such as[33, 34], and then from those reviews and previous highly cited. We summarized the functions, use cases, entities and try to figure out the role of DT. Then we search for 'digital twins' since 2019 and expend the search by snowballing reviews. By summarizing these research articles' entity (user entity, DT entity, device communication entity and observable manufacturing elements) in Table 2 we compared this with standards and reconsider the production process with the use case mentioned in the literature so that we can give answers to the RQs.

Table2 Major Entity, Function and Use case in manufacturing.

| Function Describe | Major Entity | Use case |
|---|---|---|
| Cases based on ISO/DIS 23247[35] | User entity+ Digital twin entity+ Device communication entity | |

| | + observable manufacturing elements | |
|---|---|---|
| IoT with digital twins[36] | User entity+ Digital twin entity+ device communication entity + observable manufacturing elements | |
| By using ultra-high fidelity simulation combined with history data about vehicles' health system and materials' test[37]. | Digital twin entity+ Observable manufacturing elements | |
| Structural and development process integrated by releasing computational and experimental method[38]. | Digital twin entity+ device communication entity + observable manufacturing elements | |
| Reengineering life process prediction of structure with digital twins[39] | Digital twin entity + device communication entity + observable manufacturing elements | By applying Multiphysics and multiscale damage modeling in aircraft structural life. The model has given finest analysis scale and highest fidelity models to decrease the uncertainty. While using the high-performance-computing with database share the information from simulation to design and service. |
| Spacecraft structural health monitor with uncertainty which calculating by information entropy and relative entropy in a real time prognosis and decision support system[40]. | Digital twin entity+ device communication entity + observable manufacturing elements | First by using static and dynamic analysis in part of airframe to get the critical position which can lead to fatigue cracks then propagate to failure. Measuring the crack length to update monitor model by dynamic Bayesian network. So that the uncertainty of crack extension and the data which has been measured during the whole process can be analyzed. So that the uncertainty can be further reduced. |
| A general mathematical | Digital twin entity+ device communication | To address uncertainty in the iterative multi-stage design process, and to |

| method for modeling and update data-driven design decisions[41]. | entity + observable manufacturing elements | determine the optimal fiber orientation and component thickness for composite traction (fiber steering) plane components, information from different stages of the product lifecycle was extracted. A process was devised using coupon-level experiments on fixed-wing aircraft to collect measurement data on material properties. This process involved incorporating process documentation information from manufacturing timestamps and operational loads sensed by sensors. Data assimilation was performed on the distribution and evolution of uncertain input variables. Utilizing the posterior distribution, decisions regarding the deployment design for the next stage were made based on the aforementioned information. |
| --- | --- | --- |
| Remanufacturing decision-making using product digital twins[42] | | |
| Locomotive parking brake digital twins and catalog the fault type classification[43]. | Digital twin entity+ Observable manufacturing elements | Applied parking brake then released it by using piezoelectric absolute pressure transducers measure the pressure before and after. Verify this process in DT and adjust the parameters inside the model so that the system can reflect different state of parking brake system. Based on these parameters use the internal variables to individual digital twin to train a DNN network to predict fault diagnosis. |
| Digital twins with virtual reality[44] | Digital twin entity+ device communication entity + observable manufacturing elements | Focusing on manual work station with a robotic arm to assemble children's bike. The DT offer a digital prototype for the ergonomics, safety, and robot behaviors assessments. With the verification of stakeholders, the work station will start production under the simulation and partnership with DT. |
| Predictive maintenance with | Digital twin entity+ Observable | There are four stages of using DT model the PHM, firstly define components by |
| digital twins[45] | manufacturing elements | black or white box so that the model can exactly use the algorithm to collect the parameters from analog adjustment mechanisms. With these Simulated adjustment mechanism the DT can fit in the real machine behaviors. Comparing these parameters with the actual execution then reducing the deviation between actual and digital machine. With these data after operations above the simulation can calculate RUL of components during the machine operation. |
| Rapid individualized designing of automated flow-shop manufacturing system[46] | Digital twin entity+ device communication entity + observable manufacturing elements | The DT sheet material AFMS essentially is a custom design process. Based on series of predefined reference models the data transform from the actual control to distributed semi-physical simulation. By using the simulation platform the design adjust according to the load analysis and actual operation efficiency. After all the adjustment the brand new design validate production processes with dual-layer programming so that the cost and performance can be balanced. |
| A 50MW electric drive train digital twins and optimize the run up routines digital twins for synchronous motors with DOL start[47] | Digital twin entity+ device communication entity + observable manufacturing elements | By using comprehensive CAD models and the right hand side of function as the unbalancing faults to model the sensors' data and simulation result. If there is fault happened the model will predict its progress. Because of the uncertainty's influence using the Polynomial Chaos to propagate failure degradation in confidence interval. The model verify in a 50MW transmission system belonging to the test site of the Berlin Motor Factory. |
| Plug-and-Simulate by Digital Twins and AutomationML[48] | Digital twin entity+ device communication entity + observable manufacturing elements | By using Dyvisual simulate the production process state and topology and the geometric model load and process finish in the backend of the framework so that the digital environments can be synchronized. |
| A connect micro smart factory | Digital twin entity+ device communication | The DT use the 3D model to transform the material flow in a test bench factory. By |
| digital twins[49] | entity + observable manufacturing elements | using IIoT sensors, middleware and series process the DT can monitor the production process so that the decision can be supported as well. |
| Digital twins assist with assembly in aircraft manufacturing[50] | Digital twin entity+ device communication entity + observable manufacturing elements | Based on wing flap components flexible assembly process for four components firstly according to the transfer process of coordination to build dynamic feedback. Then the DT model will reflect the coordinating elements, relationship and precision control so that the cumulative error between components will stay consistent. |
| Digital twins in RMS to RPP[51] | Digital twin entity+ observable manufacturing elements | Based on simulation of DT, the machine activation and deactivation costs of RPP can be assessed. Especially when the new need enter in the system DT can reconfigured which involved activation and deactivation of machine components between workstations between. This model is used in a sandals manufacturing factory and provide two feasible solutions of RMS' RPP problem. |
| Model-based system experimental digital twins adaptive options[52] | Digital twin entity+ device communication entity + observable manufacturing elements | By modeling in an automotive airbag production system modularly the system can generate production plan to calculate the production component utilization. While the workplace environment has been modeled in BIM which can cooperated with DT. |
| Assembly line energy optimization with digital twins[53] | Digital twin entity+ device communication entity + observable manufacturing elements | By build DT with IoT and scattered event simulation the old machines can be modeled to a SimPy process. This process is consisted of states and the transform between them. Based on DT' simulation noticed that buffer size is can be optimized which affected energy consumption and throughput of the production line. |
| Micro manufacturing unit with digital twins[54] | Digital twin entity+ device communication entity + observable manufacturing elements | The Micro Manufacturing Unit is model by Solidworks then transformed by FlexSim with the virtual components programmed by Flexscript. All subassembly models are considered static, and all dynamic features |

| | | are created and tracked at the main assembly level. Although this process will reduce the model complexity, but the purpose is for the students to understand not for the production decision. Presented model includes 102 parts, 87 of which are unique. Most of the models are based on modified OEM models, remodeling OEM models to correspond to accurate measurements and modeling the remaining parts from scratch. |
|---|---|---|
| Cloud digital twins based on cyber-physical manufacturing cloud[55] | Digital twin entity+ device communication entity + observable manufacturing elements | 3D printing component motion simulation during manufacturing operations and low consumables notification. During the manufacturing process some of the components will be heated the DT can change the color to prompt. Due to the two-way communication the DT can also optimize the power use. |
| Part data integration in digital twins using mobile and cloud technologies[56] | User entity+ Digital twin entity+ device communication entity + observable manufacturing elements | By applying MES based on Android the machine collects different parts data to record the material and cutting tool usage. Associating the product data by timestamp the process data can be monitoring so that operator can get useful information. |
| Digital twins enhance in workplace combine with ergonomics[57] | User entity+ Digital twin entity+ device communication entity + observable manufacturing elements | Verify the numerical model in an automobile assembly production line. Data collected related to the motion capture system and video recorded for each cycle. The DT use the data collected based on time simulate the workplace the output data to farther verify the ergonomics indicators. If there is a risk area the product design changes should be proposed. |
| Digital twins platform with Python Flask[58, 59] | User entity+ Digital twin entity+ device communication entity + observable manufacturing elements | A DT architecture can be accessed on web which contained three layers. The data is not only safe but also flexible to the real environment. The DT model can adapt to the evolution of materials, repair and damage of a three-layer structure experimental device. |
| Semantics in | User entity+ Digital | According to the manufacturing serial |

| aircraft design digital twins[60] | twin entity+ device communication entity + observable manufacturing elements | number and based on Dassault Systèmes V6 solution, the biunivocal relation between physical individual dummy aircraft and equivalent digital counterpart can be identified and verified. |
|---|---|---|
| Digital twins with hidden Markov model[61] | User entity+ Digital twin entity+ device communication entity + observable manufacturing elements | Five grinding state propositions were defined, Monte Carlo simulations were expressed and defined using potential states, and the surface roughness of the twin model and the real model were compared through the hidden Markov model. From the horizon of real surface height the dynamic system is quite similar with the real one. |
| Model update with digital twins for rotating machinery[62] | Digital twin entity+ observable manufacturing elements | Based on the simulation and finite element analysis the DT can obtain the critical speed and unbalance response of the rotor under different working conditions. By using the PSO optimization algorithm the model can be update in best fit of rotor system for different imbalance levels and the dynamic and simulated responses can be significantly improved. |
| Thermo-mechanical-optical coupling with digital twins[63] | Digital twin entity+ observable manufacturing elements | The DT for automotive LiDAR try to update the data based on a glass substrate which includes several conventional optical and electrical components. Based on series of thermal and optical simulation DT verify and adjust the entities so that the verify of the robustness is robust. |
| Component Modeling based on reduction which can make the model rapid adapt to the uncertainty combining with Bayesian state estimation[64] | Digital twin entity+ observable manufacturing elements | Fixed-wing UAV with online structural sensor monitor the wings' damage or degradation. By defining a model library, using Bayesian methods to obtain probability functions after treating DT as discrete random variables, and passing historical data in hidden Markov models, the model can share parameters with similar models and continuously improve. With component model this method can be easily expand to other assets more complex and larger. |

In manufacturing we noticed that DT have been used in various processes. In production design by using simulation design and analysis software DT can express the work status so that the designer can adjust the model though simulation changes. During maintenance process DT will be modeled in prognostics and health management in different statuses of maintenances carried with parameters from different components DT. In workshop production while the production progress goes on DT simulate under parameter where the data import through an architecture consists of hierarchical structures. From Table 2, it can be observed that in manufacturing industry, DT primarily focuses on manufacturing elements and DT entities. Generally, the role of DT is to transform entities into simulations through physical or mathematical modeling and by using the Internet of Things (IoT). Some researchers[65] consider virtual model as the core of model during the DT simulation. Besides, the geometric model should have standards[66, 67] for DT model to model in different systems for different use cases. So that systems' knowledge can further help the decision make clearer. But there is still a gap between the physical and digital world if it is defined by 'real', there should be a clear boundary between what is real in different fields in manufacturing not only for the model also building the knowledge extraction. The purpose of DT in specific field will help researchers to build up a DT which is more useful. However, it is evident that DT in manufacturing is more concerned with equipment communication entities and observable manufacturing elements, while less research focusing on users entity.

## 2.1.2 DT in Other Industry

For specific fields such as food preservation, Digital Twins can monitor food processing by setting conditions such as temperature, humidity, gas, light, pressure, stress, and flow in the established models. However, in many food applications, specific applications and models are often scattered throughout the industry, meaning there is no universally applicable way to apply Digital Twins. [68]. In other industry, such as mineral collection[69] like open-pit mining production, the role of DT focuses on the mining sequence under different constraints, which can be understood as production scheduling[70]. Another issue of concern is the operating cost during the mining process, the main cost is the equipment operating cost. DT can use real-time modeling of closed fleet management to sense physical assets to the cognitive layer to ultimately assist decision-making. In oil reservoirs storage[71, 72], during the drilling process, the DT model continuously updates dynamic weighted predictions based on the increase in time and section depth to preventing failures. In the oil and gas industry, although the terminology used to describe assets in the industry is different from other industries, such as drilling rigs in the production process, and the components installed in oil and gas facilities are also different, DT still be structured in a similar way, from asset performance management to in risk assessment, what more important is professional domain knowledge, and DT only plays a supporting role to help people generate other insights in the knowledge. Agriculture[73] faces high quality[74] and environmental demands[75] due to its complexity and dynamics, and its dependence on many uncertain factors such as weather and seasonality. DT can simulate, optimize and make decisions. Some interesting DT[76] use sensors to build digital potatoes. Though adjusting the parameters of the harvester and work with the cloud platform to achieve price estimation and other needs.

Although these DT have different use cases but they have similarities as well. All these DT need a data support which is usually come from sensors through industrial communication like IEEE 802.11n-2009[77, 78] set inside the space by simulated or on scene before the DT is built while OPC[79] and MTConnect[80] provide standards try to integrate individual specifications into one extensible framework which can be an assistive tool used in specific traditional field like predictive maintenance.

In summary, DT is making significant strides across various industries, including manufacturing, food preservation, mineral collection, oil reservoir storage, and agriculture. In food preservation, DT monitors processing processes by observing conditions such as temperature, humidity, gas, light, pressure, stress, and flow. However, specific applications and models in the food industry are often fragmented, lacking a universal approach to implementing DT. Similarly, in mineral collection, DT implementation focuses on optimizing mining sequences under different constraints, akin to production scheduling. Concerns in this field revolve around operating costs, particularly equipment operating costs, which DT addresses through real-time modeling and closed fleet management.

By using model to monitor processes through DT has shown significant efficacy across industries. Moving forward, exploring the mathematical methodologies used in DT applications will provide further insights into potential and effectiveness.

## 2.2 Methods in DT Systems

For the overall life cycle of the manufacturing, the production system is a dynamic system[81]. Different methods can be used for different goals, and there is no universality. The problem encountered in production is often that allocating limited resources to task elements constraint with tasks time et.al[82]. The method to solve these is generally mathematical modeling of the problem[83], and by given constraints, using methods such as linear programming or mixed integer programming to find the optimal solution or local optimal solution, according to the solution system can provide advice based on data to make decision.

Although DT and mathematical model both have 'model' but DT focuses more on physical modeling which means to transform physical into the virtual and pays less attention to the mathematical form of the problem. From the introduction we notice that from start of modeling to feedback instructions through DT in a closed loop. This loop contains different parts can build up a system. The goal of system is different due to the scenes and the mission[84].

Here, in order to find the distance between theory and implementation of specific DT, we try to enter from a mathematical perspective and analyzes the theoretical issues that may be involved. Less researches focused on mathematical modeling of DT. Chinesta et.al proposed Hybrid Twins[85], which describe how to eliminate these intricate correlations to visualize reality. We search in Google scholar by using the key words of 'mathematical model DT'. Although there are 16,900 results from 2020, but only few focused on the mathematical model in DT[86]. The modeling process will not only involve the change of the system state over time in the specified domain but also the defined virtual state. For a specific component or part, high-dimensional design and experimentation are indispensable, in addition to low-dimensional specific operational steps and corresponding data perception. The model explanations in physical and digital are both important for analyzing even more predicting with the model. The results of model explanations can help understand the decision-making process from the designer's perspective[87].

## 2.2.1 State Driven Model

### Probabilistic Model

Probabilistic Digital Twins (PDT)[88] can model the parameters and states of physical systems applied in the environment and utilize Bayesian networks for dynamic updates. Due to the uncertainty and dynamic updates within DT, PDT are well-suited for dynamic assessment of DT systems.[89]. It can model the parameters and states of physical systems applied in the environment and utilize Bayesian networks for dynamic updates. Modeling in such scenarios emphasizes the dynamics of time with environment. Some scholars argue that models relying solely on physical principles for discrete events in DT are vulnerable to data noise which has led to the adoption of surrogate models to quantify uncertainty probabilistically and infer function distributions[90]. Moreover, they improve models by creating single-degree-of-freedom models on a time scale, highlighting the significance of time and distinguishing with the concept 'slow time' to evaluate system uncertainty[91]. Furthermore, the integration of expert models with multiple time scales allows for more accurate tracking of parameter scale evolution over time[92]. Markov chain optimization combined with Bayesian theory is occasionally employed in DT system design[87]. Dimensionality reduction and reward functions, represented by states, can be utilized to account for dynamic changes as the environment evolves from a design perspective.

The Markov model's characteristic of continuous events in DT relates the probability of an event occurring at the current moment solely to the event at the previous moment[93]. This feature helps comprehension of the mathematical modeling process in DT. Utilizing statistical methods like Bayesian analysis[94, 95], the dynamic system can be abstracted into a probabilistic graphical model, offering flexibility over time[96]. They assume $t$ represents the observation time during DT, $S$ and $D$ respectively denote the random variables representing the physical state and the digital space evolving over time, and the data observed flowing into the digital space is defined as $o$.

Because the parameters in the digital space are complex, they uses an unspecified independent structure. The advantage of doing so is that it allows for the decomposition of variables with joint distributions, enabling each state to be reconstituted based on observation. The Bayesian probability formula for the defined graphical model is as follows:

$$p(D_0, \ldots, D_{t_c}, Q_0, \ldots, Q_{t_c}, R_0, \ldots, R_{t_c} | O_0, \ldots, O_{t_c}, \mu_0, \ldots, \mu_{t_c})$$

$$\phi_t^{update} = p(D_t | D_{t-1}, U_{t-1} = \mu_{t-1}, O_t = o_t)$$

$$= \int p(Z_t | Z_{t-1}, e, U_{t-1} = \mu_{t-1}, O_t = o_t) p(e) \, de$$

$$\phi_t^{QoI} = p(Q_t | D_t)$$

$$\phi_t^{evaluation} = p(R_t | D_t, Q_t, U_t = \mu_t, O_t = o_t)$$

$$p = \prod_{t=0}^{t_c} [\phi_t^{update} \phi_t^{QoI} \phi_t^{evaluation}]$$

Bayesian priori defined numerical space state is represented by discretization in finite element space. As a second-order ordinary differential equation of force on the finite element grid with respect to mass, damping, and stiffness over time, formulated as follows:

$$M(d)\ddot{x}(t) + V(d)\dot{x}(t) + K(d)x(t) = f(t)$$

Where $M$ represents mass, $V$ represents damping, and $K$ represents stiffness.

In the process of Bayesian estimation, [96]in the above probability distribution, the precise geometric parameters of the structure measured in the coupon level experiments are used. The rewards function is based on the DT input control. A weighted sum of the reduction percentages in the initially defined series of material stiffness is calculated and selected through a reward mechanism to determine suitable strategies. The reward is defined as follows:

$$r_t^{health}(q_t) = \frac{\epsilon_{max} - max_j(\epsilon_t^j)}{\epsilon_{max}} [96]$$

Where $\epsilon_{max}$ represents the maximum allowable strain level.

The estimation of geometric parameters defined in experiments is:

$$g = [l, c]$$

Where $g$ represents geometric parameters included in the experiment, while $l$ and $c$ are geometric parameters of the experimental objects defined by author for specific scenario.

They calculate uncertainty using incremental updates, obtaining a posterior distribution with uniform weights via an iterative state estimation method like standard particle filtering. They approximate a density-weighted credible interval, transform it by inverse sampling through the empirical cumulative distribution function, and compute and propagate uncertainty. They correct uncertainty by conditionally independently decomposing DT update models, known as 'data assimilation':

$$\phi_t^{assimilation} = p(O_t = o_t | Z_t) = \int p(O_t = o_t | Z_t, e) \, p(e) de$$

During prediction of DT, the issue shifts into optimizing the DT within a given range to find the optimal strategy. Here, the state comprises combinations selected from the digital space and the graphical probability. The reward function of MDP process is by using a weighted sum, as follows:

$$r_t = r_t^{health}(q_t + \omega r_t^{control}(\mu_t))$$

In addressing uncertainty in DT using Bayesian state quantification, some scholars integrate physics and data using hybrid twins. They employ discrete partial differential equations to elucidate the discrete physical model actions in DT. After reducing the model order, they utilize difference equations along with Bayesian rules to clarify the mathematical relationships of variables over discrete time, while also quantifying errors and uncertainties[64].

Apart from sequential decision-making above, DT can also be used in material characterization to analyze the effects of processes during material characterization. To reduce inevitable errors and the dimensionality, some scholars use manifold structures to express observed data[97]. Although the idea of using kernel functions like SVM to map data space to low dimensions is similar to manifold learning, this approach appears unstable when faced with large datasets[98]. The author introduced the method of manifold learning data when analyzing the performance of composite materials, such as using PCA and diffusion map to reduce the dimensionality of independently distributed original experimental data, to reducing the computational burden.

They define the dimensionality reduction data $[H]$ as a random matrix $[\eta]$, and defines the obtained non-parametric Gaussian mixture model as:

$$p[H]([\eta]) = pH(\eta^1) \times \ldots \times pH(\eta^N)$$

To explore the potential joint intrinsic structure of the data points, they assumed that the original data consists of independent standard Gaussians $[N]$, the reduced random matrix after dimensionality reduction $[Hd]$, and the diagonal matrix $dW(r)$ obtained from the normalized diffusion map. Since the coefficients in the second-order stochastic differential equation for $Y(r)$ are only functions of $Xt$, this process can be regarded as an Ito process[99, 100].

$$d[Y(r)] = [L([Z(r)][g]^T)][a]dr - \frac{1}{2}f_0[Y(r)]dr + \sqrt{f_0}[dW(r)][a]$$

Solving this Ito process can obtain the state of the stochastic process at a given time point. By utilizing the numerical method of implicit midpoint, problem solved[101]. They employed manifold learning twice to acquire the reduced form of the original dataset and inferred that this form conforms to a non-singular matrix form of the Ito process. In PDT, by utilizing probabilistic learning on manifolds, computational complexity in large databases can be reduced. This gives PDT an advantage in problems such as composite material performance analysis. However, it is equally important to emphasize that whether in PDT or DT, the use of different numerical analysis methods requires a detailed description of specific scenarios and the determination of variables involved in uncertainty.

## Virtual Sensing

DT systems rely on physical information to model and update the system. DT can be given senses by using physical sensors to transmit or fuse data. However, incorporating all physical information and information between components is difficult for DT modeling. This is not only because deploying sensors can be expensive and they are environmentally or technologically unfeasible[102, 103]. Beyond this, physical sensor drift is inevitable. This allows virtual sensors to improve.

In general, virtual sensors can be only based on physical sensors or mixed based on physical and virtual sensors[104]. It can introduce the solved parameters into the physical model, and under the premise of following physical laws the model finds the relationship between variables, which helps to explain the DT system, thereby enabling DT to complete specific tasks.

Here[105], the role of virtual sensors is to model the process variables that cannot be physically represented and cannot be detected in multi-component DT. Introduce the quasi-static conservation equation[106] through the conservation law. The function of this equation is to serve as a constraint in the process variables. By solving this equilibrium equation under the condition of incompressible flow with a single inlet and outlet, the values of the virtual sensor with respect to flow, pressure and temperature are obtained. Then bring it into the multi-component model to generate model residuals for fault diagnosis[107].

Through the literature review, we found that virtual sensors are more commonly used in the construction industry. The use of virtual sensors for building operation and maintenance processes in the construction industry can be a great way to save operating costs. Moreover, for the system as a whole, the act of installing physical sensors may cause leakage or failure of the system. In addition to providing intended functions, devices within buildings can also serve as virtual sensors themselves [108]。

Virtual sensors in the construction industry are also based on physical laws to model mass, temperature or pressure, such as in vapor compression systems[109], for the electronic expansion valve (EXV) used to control refrigerant flow and pressure, it is difficult to indicate the current opening of the EXV. Because this requires designing the number of moving steps at the initial position, but it will inevitably produce errors, so in order to determine the relationship between the two-phase inlet and EXV, the PI theorem[110]. Treat the valve as a virtual sensor, and find the data points used to establish correlation on the curve based on the mixing curve obtained from the expansion process of the inlet and outlet states of the main valve and secondary valve. Then according to[111], these data are divided into 17 dimensionless PI groups, and the continuous correlation for individual and general is finally obtained from the independent variable-dependent variable relationship.

H. Edtmayer et al.[112] established a DT using real-time measurements of indoor space on a self-built physical testing facility. On this basis, they placed virtual sensors and combined computational fluid dynamics to obtain the boundary conditions of the model. Thus, the thermal comfort of the indoor space of the physical testing facility was finally found accurately.

In these specific scenarios, the location and scale of the target model where the virtual sensor is located are extended to the space and time of the model. At the same time, data-driven virtual sensors that minimize residual errors can allow the model to take greater advantage in time distribution.

Virtual sensors should maintain the same level of performance as physical sensors. DT serves as a simulation carrier for the process. With the help of virtual sensors, the system can be made more complete and robust.

## 2.2.3 Data Driven DT

### Rule-based DT

Rule-based DT generally uses DT as a carrier to gather specific domains knowledge. Knowledge graph is a tool that can describe dynamic relationships between data[113]. Key content can be extracted by using knowledge graph. In DT, data not only consists of dynamic information but also includes static data about material assets. Utilizing the meta-model of knowledge graph for data transmission, enables internal linking and knowledge completion within DT, thus further facilitating querying and inference in both dynamic and static data[114].

Moreover, Finite State Machines (FSM) has been utilized by scholars to detect potential attacks within DT system[115], due to its inclusion of the states within a system, the input-output relationships, and transitions between states[116]. Specifically, after defining $P := (X, x_0, U, Y, \delta, \lambda)$, external inputs are referred to as stimuli $S$.

$$S := \{z \in \widehat{U} | z \in U \wedge z \notin Y^*\}$$

If the observation of two identical stimuli in the physical environment results in DT performing the same actions in different environments, it is necessary to reevaluate the logic and examine the data sources and objectives.

Furthermore, semantic model can work alongside machine learning to support event recognition and decision-making in smart city scenarios, such as building energy analysis [117]. Moreover, belief rule base[118] by using the white-box model to identify the association rules among parameters in DT system. The contribution of parameters calculate by rules can be a reference with the assistance of expert knowledge.[118]. Based on the parameter count $M$, active rules $R_k$, confidence distribution $\beta_n$, and final output $y_p$ are defined as the parameters to be controlled in the DT. The four variables defined are sequentially used to calculate matching degree $ctr(I_m^p, R_k)$, contribution $ctr(R_k, \beta_n)$, and utility $ctr(\beta_n, y_p)$. The contribution $ctr(R_k, \beta_n)$ is obtained by taking the partial derivative of the rule weight $\omega_k$ and distribution $\beta_{n,k}$ using evolutionary algorithm. These three $ctr$ multiplied to determine the primary parameters influencing the output $y_p$. This process enables the identification of the primary parameters influencing $yp$, starting from the desired control of $yp$.

## Machine Learning and DT

Machine learning (ML) can classify data based on labels or dataset and make predictions based on datasets in various scenarios. Due to these powerful capabilities, ML serves as a potent tool across disciplines to address problems. Here, we start with the current capabilities of machine learning and analyze the opportunities and challenges that machine learning in the context of DT may bring to the DT.

As we approach it from the perspective of machine learning, we simply categorize Machine learning (ML), Deep Learning (DL), and Reinforcement Learning (RL). The intention behind this classification is based on their ease of use and data characteristics. For instance, ML methods like XGBoost[119] have an advantage in handling highly correlated data, while DL and RL don't. Similarly, DL, particularly using Convolutional Neural Networks (CNN[12]) as the foundation, such as ResNet[120], is advantageous in processing natural data types like images, speech, and natural language process[121], especially for classification tasks when labels and processing are well-defined. CNNs can swiftly and accurately perform classification or other related tasks.

However, for tasks requiring network reward and punishment or tasks where networks or agents need to accomplish goals according to certain rules, RL methods like Deep Q-Networks (DQN)[122]. ML may use multiple networks to handle different categories of inputs, using methods like federated learning[123], to address specific tasks in the scenario. Similar to DT, ML as a powerful tool, still requires emphasis on the specific circumstances and datasets it utilizes.

Although some research[124, 125] has begun to integrate DT and ML, the focus of these studies tends to lean more towards the applications of ML rather than on the DT's physical or mathematical modeling.

The emergence of model reduction techniques can assist DT developers in safely compressing DT models without losing crucial information[126]. Methods for model reduction include Krylov subspace methods[127]and one-dimensional Burgers equation[86]. Based on the results of model reduction, components of the composite model can be decomposed using system design methods and monitored according to system requirements and component properties. Similarly, the advent of model reduction allows for the subdivision of DT or PDT into finer and quantifiable components, enabling machine learning to better operate within the reduced models.

Moreover, viewing each component as an independent entity, ML can be leveraged to predict overall model behavior and possible operational outcomes. By using federated learning, data from each component can updated on separate clients, and the aggregated parameters adjusted using joint averaging algorithms. The goal is to minimize local loss function and distributing the aggregated parameters to each component iteratively. While such methods rely on the ability of all components to communicate, they provide a different perspective for complex DT systems combined with ML. In DT systems, the utilization of federated learning inevitably encounters resource constraints. Some scholars argue that in the IoT context, reward functions for DT may become unstable due to DT's errors[128]. To address this concern, they incorporate DQN into federated learning. While DQN is a common reinforcement learning method suitable for learning and adjusting states based on rewards within a given environment, they adapt it in this scenario to allocate and aggregate device resources in DT based on federated learning node capabilities.

Similarly, due to DQN's learning capabilities, it can also be applied to mathematical problems that require step-by-step resolution. Some scholars utilize DQN for sequential decision-making in PDT[89]. The states of DQN can be understood as random states of PDT in discrete time. Selecting the optimal strategy within a given environment becomes a matter of how to use DQN to enable PDT to choose the best solution in dynamic programming problems.

The use of ML in DT or PDT is becoming increasingly popular. We believe with the continuous development of neural networks to solve different problems, there will always be suitable networks to ultimately address specific issues in particular scenarios. For the DT industry, the surge in black-box ML algorithms has brought challenges. These algorithms involve numerous parameters, making it difficult for DT practitioners to explain which may lead to misunderstanding of results. To address such challenges, using interpretable machine learning techniques like SHAP[129] and LIME[130] could be a good choice. However, it's better to encourage DT industry to have a clear understanding of these black-box algorithms. Meanwhile, although ML methods can be used to address issues in DT, such as uncertainty analysis, the cost of data collection for ML is beyond expensive.

# 3 Discussion

With the continuous development of digitalization, Digital Twins (DT) will see ongoing new applications and trends across various industries. If considered DT as a service to business or a tool within specific industries, it can assist industries in transitioning from simulation to control, achieving full lifecycle management. In manufacturing, DT can help solve problems such as production monitoring, analytical decision-making, and predictive maintenance. Similarly, in other sectors such as petrochemicals and agriculture, DT can adjust implementation plans flexibly to accomplish specific tasks.

The goal of these jobs usually is updating dynamically or calculating elements. Especially in some cases the DT can give operators inner sight of black box. But for some decision-making system which DT can play roles needs horizons not only for only one part of the system. In those cases, boundaries between physical world and DT should be clearer. So does the boundary between DT with other modules of whole system.

Furthermore, Comparing the efficacy of DT implementations concurrently is challenging due to varying industry-specific demands for accuracy and precision in different scenarios. Meanwhile, the cooperation among elements within DT is important for the holistic functioning of the system. These dual aspects underscore the interconnection among simulation, algorithms, and data across diverse domains and contexts.

Despite the continuous development of digitalization, the journey of DT from conceptualization to practical implementation and tangible benefits for people's lives still has a distance to go. While DT has gradually matured in its development trends since its inception, its concrete application and real-world impact are still evolving. With the development of related technology such as decentralization[131] and cloud manufacturing[132], DT may transform from huge systems into independent and autonomous small systems[133]. Each system will be based on its assigned tasks and location. The location here is not only It is the geographical location, but also includes the roles it plays in different collaboration tasks.

The complexity of DT modeling poses a significant challenge. Whether through physical modeling or the integration of methods like PDT with numerical analysis such as Bayesian inference or model reduction, DT demands specific scenarios and conditions—there is no one-size-fits-all approach. However, in specific scenarios like fault diagnosis or planning decisions, leveraging DT combined with machine learning and other artificial intelligence methods can provide a more comprehensive analysis of specific issues. Therefore, incorporating various methods into DT might prove beneficial.

Looking ahead, DT will encounter a wide variety of data types, necessitating not only the ability to quantify uncertainty but also the capability to fuse data[134]. As DT continues to evolve, addressing these challenges and harnessing its full potential will require ongoing research, development, and practical implementation efforts. Thus, while DT has made significant strides, its journey towards becoming an integral part of everyday life and enhancing human experiences is still unfolding.